\documentclass[12pt,preprint]{aastex}

\title{The Large Scale Cosmic-Ray Anisotropy as Observed with Milagro}

\shorttitle{Cosmic Ray Anisotropy}

\author{A. A. Abdo\altaffilmark{1,2}, B. T. Allen\altaffilmark{3}, 
T. Aune\altaffilmark{4}, D. Berley\altaffilmark{5}, S. Casanova\altaffilmark{6}, 
C. Chen\altaffilmark{3}, B. L. Dingus\altaffilmark{6}, \\ R. W. Ellsworth\altaffilmark{7}, L. Fleysher\altaffilmark{8}, 
R. Fleysher\altaffilmark{8}, M. M. Gonzalez\altaffilmark{9}, 
J. A. Goodman\altaffilmark{5},\\ C. M. Hoffman\altaffilmark{6}, B. Hopper\altaffilmark{5}, P. H. H\"{u}ntemeyer\altaffilmark{6}, B. E. Kolterman\altaffilmark{8}, C. P. Lansdell\altaffilmark{5},\\ J. T. Linnemann\altaffilmark{10}, J. E. McEnery\altaffilmark{11}, A. I. Mincer\altaffilmark{8}, P. Nemethy\altaffilmark{8}, D. Noyes\altaffilmark{5}, J. Pretz\altaffilmark{6},\\ J. M. Ryan\altaffilmark{12}, P. M. Saz Parkinson\altaffilmark{4}, A. Shoup\altaffilmark{13}, G. Sinnis\altaffilmark{6}, A. J. Smith\altaffilmark{5}, G. W. Sullivan\altaffilmark{5},\\ V. Vasileiou\altaffilmark{5}, G. P. Walker\altaffilmark{6}, D. A. Williams\altaffilmark{4}, and G. B. Yodh\altaffilmark{3}}

\shortauthors{A. A. Abdo and et al}

\altaffiltext{1}{National Research Council Research Associate, National Academy of Sciences, Washington, DC 20001}
\altaffiltext{2}{Space Science Division, Naval Research Laboratory, Washington, DC 20375}
\altaffiltext{3}{Department of Physics and Astronomy, University of California, Irvine, CA 92697}
\altaffiltext{4}{Santa Cruz Institute for Particle Physics, University of California, 1156 High Street, Santa Cruz, CA 95064}
\altaffiltext{5}{Department of Physics, University of Maryland, College Park, MD 20742}
\altaffiltext{6}{Group P-23, Los Alamos National Laboratory, P.O. Box 1663, Los Alamos, NM 87545}
\altaffiltext{7}{ Department of Physics and Astronomy, George Mason University, 4400 University Drive, Fairfax, VA 22030}
\altaffiltext{8}{Department of Physics, New York University, 4 Washington Place, New York, NY 10003}
\altaffiltext{9}{ Instituto de Astronom\'{i}a, Universidad Nacional Aut\'{o}noma
 de M\'{e}xico, D.F., M\'{e}xico, 04510}
\altaffiltext{10}{Department of Physics and Astronomy, Michigan State University,
 3245 BioMedical Physical Sciences Building, East Lansing, MI 48824}
\altaffiltext{11}{NASA Goddard Space Flight Center, Greenbelt, MD 20771}
\altaffiltext{12}{Department of Physics, University of New Hampshire, Morse Hall, Durham, NH 03824}
\altaffiltext{13}{The Ohio State University, Lima, OH 45804}

\begin{abstract}Results are presented of a harmonic analysis of the large scale
cosmic-ray anisotropy as observed by the Milagro observatory. We show a 
two-dimensional display of the sidereal anisotropy projections in right 
ascension generated by the fitting 
of three harmonics to 18 separate declination bands. The Milagro
observatory is a water Cherenkov detector located in the Jemez mountains 
near Los Alamos, New Mexico. With a high duty cycle
and large field-of-view, Milagro is an excellent instrument for measuring
this anisotropy with high sensitivity at TeV energies. The analysis is 
conducted using a seven year data sample consisting of more than 95 billion 
events, the largest such data set in existence. We observe an anisotropy with 
a magnitude around 0.1\% for cosmic rays with a median energy of 6 TeV. The 
dominant feature is a deficit region of depth (2.49 $\pm$ 0.02 stat.\ $\pm$ 
0.09 sys.) $\times 10^{-3}$ in the direction of the Galactic North Pole 
centered at 189 degrees right ascension. We observe a steady increase in the 
magnitude of the signal over seven years. 
\end{abstract}


\begin{document}
\maketitle

\section{Introduction}

Observation of the sidereal large scale cosmic-ray (CR) anisotropy at
energies of 1 - 100 TeV is a useful tool in probing the magnetic field
structure in our interstellar neighborhood as well as the distribution of
sources. Cosmic-rays at these energies are almost entirely of Galactic origin 
and are expected to be nearly isotropic
due to interactions with the Galactic magnetic field (GMF) \citep{AsCR}.
The gyro radii of CRs at these energies in a GMF of about $1\mu$G are from
about 100 AU - 0.1 pc which is much smaller than the size of the Galaxy. This
has the effect of trapping the CRs in the Galaxy for times on the order of
$10^{6}$ years. Inhomogeneities in the GMF are interspersed randomly
throughout the galaxy and in effect act as
scattering centers for CRs. This randomizes their directions as they propagate
leading to a high degree of isotropy on scales of a few hundred
parsecs.

Anisotropy can be induced through both large scale and local magnetic
field configurations which cause deviations from the isotropic diffusion
approximation. At lower energies of around a TeV, the heliosphere
may be able to induce a CR excess in the direction of the heliotail
and also could modulate the overall CR anisotropy \citep{NgLC,NgTI}. At higher
energies, the contribution of discrete CR sources has been shown to be capable
of creating a large scale anisotropy \citep{PtSNR,SMP}.

Diffusion of CRs out of the Galactic halo can also produce an anisotropy. Since
the matter density is higher in the Galactic disk compared to that in
the surrounding halo, the diffusion coefficient will generally be much higher
in the halo. For this reason, CRs produced in the Galactic disk will tend
to diffuse out into the halo creating an anisotropy in the direction
perpendicular to the disk. Predictions of the anisotropy have been made
using values of the diffusion coefficient inferred using a given propagation model and 
observational data. This predicted anisotropy can take on values
of $10^{-2}$ to $10^{-5}$ depending on the propagation model used \citep{AsCR}. Given this correlation between the anisotropy and
diffusion coefficient, knowledge of the large scale CR anisotropy can be used
to constrain diffusion models.

In addition to the above effects, Compton and Getting introduced a
theory \citep{CG} of CR anisotropy which predicts a dipole effect due to the
motion of an observer
with respect to an isotropic CR plasma rest frame. The anisotropy arising
from the Earth's motion around the sun is calculated to be on the order
of $10^{-4}$ and would appear in universal time. This effect has been
observed by numerous experiments (e.g. see \cite{ET}, \cite{TBCG}). There is also a possible sidereal time
anisotropy coming from the motion of the Solar System through the
Galaxy. This effect is more difficult to predict given that the isotropic
CR rest frame is not known. Using the assumption
that the CR rest frame does not co-rotate with the Galaxy leads to a predicted
anisotropy on the order of $\sim0.1\%$. This effect has not
been observed \citep{TB}.

There have been numerous observations of the large-scale sidereal anisotropy 
in the range of energies $10^{11} - 10^{14}$ eV. In this paper we use large-scale 
anisotropy to refer to features of the anisotropy in the sky with an extent greater than 
$\sim 40^{\circ}$ in right ascension. Most of these observations are examined by fitting
harmonics to the distribution of cosmic-ray events in the right ascension
direction. The main features of the anisotropy, the amplitude and phase, are 
fairly constant over the energy range mentioned above ($3-10 \times 10^{-4}$ 
for the amplitude, and $0 - 4$ hr for the phase, see \cite{SK} for a 
compilation). At TeV energies, the focus of this paper, the anisotropy
is known to have a value on the order of $10^{-3}$ with a deficit region,
sometimes called the ``loss-cone", around $200^{\circ}$ right 
ascension, and an excess, or ``tail-in" region, around $75^{\circ}$ \citep{NgLC,NgTI}. 
In this paper, we present the results of a harmonic analysis of the large scale 
cosmic-ray anisotropy in the northern hemisphere as observed by the Milagro experiment.

\section{The Milagro Observatory}

The Milagro observatory \citep{Milagro} is a water Cherenkov detector which 
is used to monitor extensive air showers produced by TeV gamma-rays and 
hadrons hitting the Earth's atmosphere. Milagro is located in New Mexico at a
latitude of $35.88^{\circ}$, a longitude of $106.68^{\circ}$, and an altitude of 2630 m above sea level,
possessing a large field of view 
of $\sim$2sr and a high duty factor of $>90$\%. The detector is composed of
an 80m$\times$60m$\times$8m pond filled with $\sim 23$ million liters of
purified water and protected by a light-tight cover. The central 
pond is instrumented with two layers of PMTs: a top layer with 450 PMTs
under 1.4m of water which detects Cherenkov light from air shower 
electrons, electrons Compton scattered by gamma-rays, and gamma-rays that have 
converted to electron-positron pairs in the water; and a bottom layer with 273 
PMTs 6m under the surface used for gamma-hadron separation (not used in
this analysis). The direction of an air shower is reconstructed 
using the relative timing of the PMTs hit in the top layer of the pond with
an angular resolution of $<1^{\circ}$.

This pond is surrounded by a 200m$\times$200m array of 175 ``outrigger"
tanks. Each ``outrigger" is a cylindrical, polyethylene tank with a
diameter of 2.4 m and a height 
of 1 m. The outrigger tanks contain $\sim 4000$ liters of water and are 
instrumented with a single downward facing PMT located at the top of the 
tank. The inside of each tank is also lined with Tyvek (a white, reflective
material) to increase the light collection capability of the 
PMT.  The outrigger array, completed in 2003, is used to improve the angular
resolution and in the estimation of primary particle energy.

The data used in this analysis were collected by Milagro from
July 2000 through July 2007. The hardware trigger was essentially a voter
coincidence, requiring a multiplicity of about 75 PMTs. During this time 
the trigger multiplicity was tuned to maintain an average trigger rate of $\sim$1700 events per second 
(to within 10 \%), the majority of which are due to hadronic showers. After event reconstruction, 
accepted events are required to have used at least 50 PMTs in the angular fit 
($N_{fit}$) and have a zenith arrival angle of $\leq50^{\circ}$. The $N_{fit} \geq 50$ requirement 
removes effects due to fluctuations in the trigger 
threshold as events passing this cut generally have a higher PMT 
multiplicity than the hardware trigger requires. Although this makes the data more stable, it also 
reduces the effective trigger rate to $\sim 500$ Hz. After these cuts the 
data set consists of $9.59\times10^{10}$ cosmic-ray events with a median 
energy of 6 TeV as determined from the measured primary cosmic-ray spectra 
and simulated detector efficiency. The median energy varies slowly with zenith 
angle; it is 4 TeV for zenith angles from $5^{\circ}$ to 
$10^{\circ}$, and 7 TeV for zenith angles from $35^{\circ}$ to 
$40^{\circ}$.

Detector response to primaries is determined with a Monte Carlo simulation. This 
Monte Carlo has two main components: the simulation of an air shower in the 
atmosphere and the simulation of the shower through the Milagro detector. These 
simulations are carried out using the CORSIKA \citep{CSKA} and 
GEANT4 \citep{GEANT} packages respectively. Cosmic-ray nuclei ranging in 
mass from the dominant protons to those as heavy as iron
are simulated according to their spectra as measured by the ATIC 
experiment \citep{ATIC}. Because of the large shower to shower variations, Milagro 
cannot measure the energy of individual primaries generating these air showers without 
careful data selection criteria. Measurement 
of energy correlated parameters does allow some determination of primary energy 
spectra without severely restricting the data. The energy analysis of the large-scale 
anisotropy will be the subject of a future publication.

\section{Data Analysis}

\subsection{Forward-Backward Asymmetry and Harmonic Method}
\label{fbahm}

\begin{figure}
\epsscale{.80}
\plotone{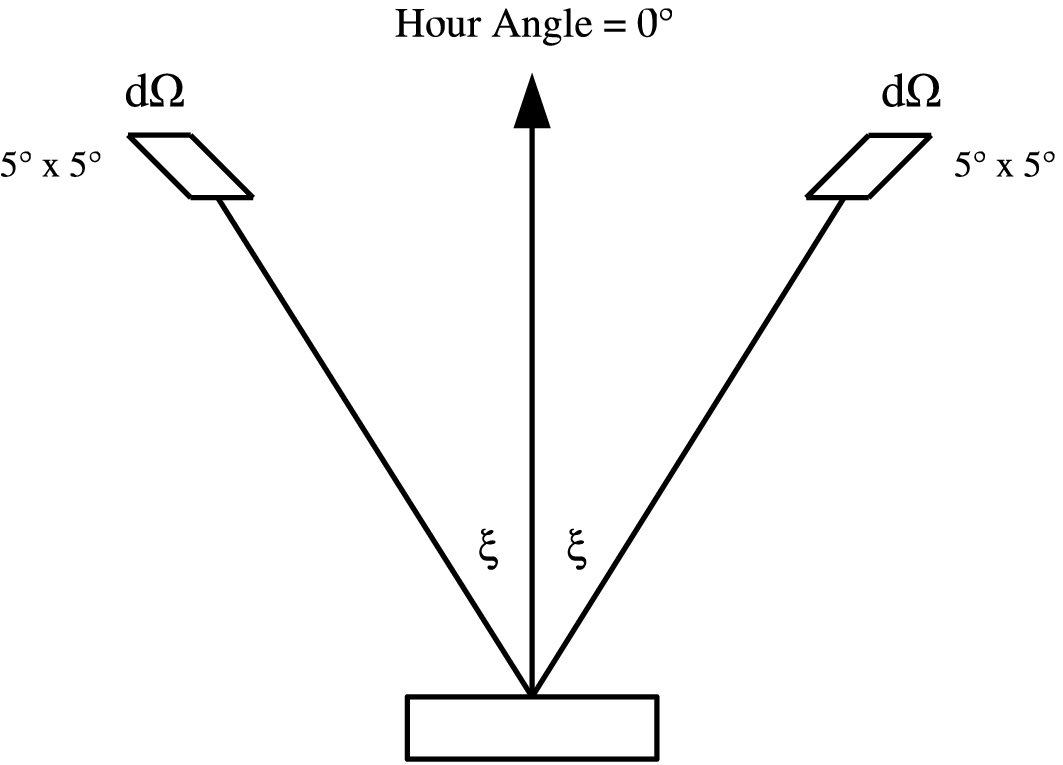}
\caption{Diagram showing the definition of $\xi$ used in the calculation of the
forward-backward asymmetry for a single dec.\ band and a given 30 minute 
histogram. $\xi$ is in the direction of hour angle. See text in Section 3.1 for
details.}\label{fig1}
\end{figure}

We have designed an analysis method to search for fractional deviations
from isotropy down to levels of about $10^{-4}$, below the level of
predictions of the Compton-Getting effect and the magnitude of previous
observations. Such sensitivity cannot be achieved by looking at
raw event rates as a function of direction, because the air shower development
essentially makes the atmosphere part of the detector, and weather effects
lead to typical overall trigger rate variations of $\pm 15 \%$. Thus, counting 
raw event rates would give random apparent anisotropies
completely overshadowing a true signal. Instead, we employ the technique of
forward-backward asymmetry (FB) using the number of events ($N_{F}$
and $N_{B}$) collected in some small time interval in two ``telescopes", i.e.\
pixels of small and equal solid angle, at the same forward and backward
angle as shown in Figure \ref{fig1}.

\begin{equation}
FB = (N_{F} - N_{B})/(N_{F} + N_{B})
\end{equation}

The expression of FB is manifestly independent of overall
detector rate, as can be seen from the invariance of FB under the
substitution $N_{F}, N_{B} \rightarrow cN_{F}, cN_{B}$, where $c$ is
a constant.

 The analysis method utilizes the rotation of the Earth to search for a
coherent modulation of FB during a 24 hour day. The FB modulation does
not directly give the anisotropy of the sky, but is a function of it. As
a region of the sky with an excess cosmic-ray flux relative to the average
(a positive anisotropy region) is swept into the forward ``telescope" FB
becomes more positive; when the same excess is swept into the backward
``telescope", FB becomes negative. One can think of FB as a coarse
``derivative" of the actual anisotropy of the sky. Thus the FB modulation
is a tool to obtain the quantity of interest, the fractional anisotropy
(i.e.\ deviation from uniformity) of the sky, as detailed below.

The FB method is commonly used in particle physics (e.g.\ see \cite{FBCA})
and is also closely related to the East-West technique, classically 
used in other anisotropy measurements (for an example see \cite{NgLC}), in 
which the ratio examined for modulation is 
$EW = (N_{E} - N_{W})/N_{tot}$, formed with the rate integrated over the 
whole Eastern and Western sky. Again this is a rate independent quantity which
is used to remove random fluctuations in overall detector rate, a common
requirement in all experiments using the atmosphere as part of their
detector. The EW method can be thought of as a single integrated
measurement, compared to 
our multiple localized and independent measurements of FB which are
used to improve the statistical power of this analysis.

\begin{figure}
\epsscale{.80}
\plotone{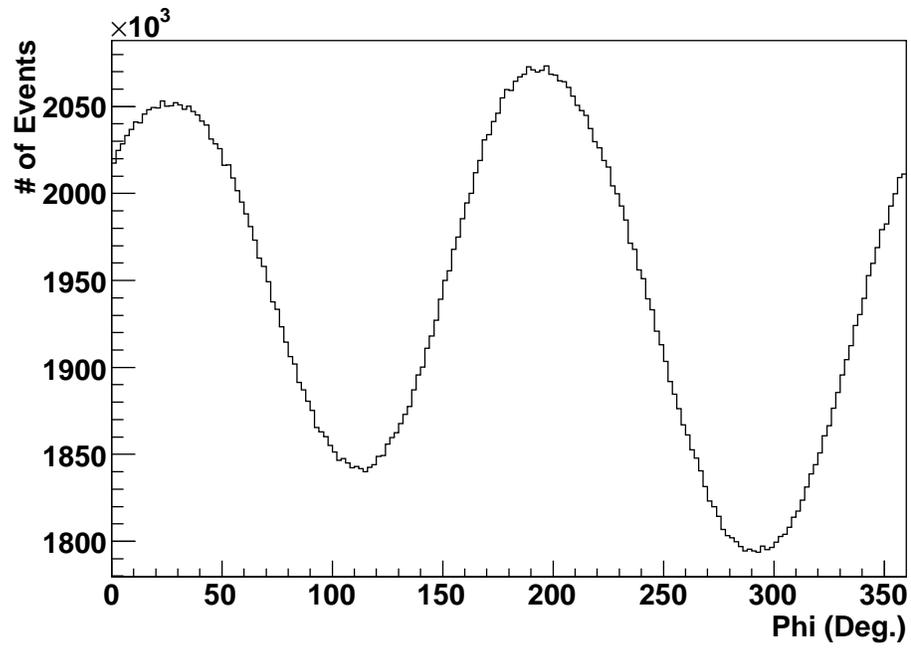}
\caption{Representative plot of the number of events collected vs.
azimuthal angle for the Milagro detector over the course of three days. The
observed variation is the inherent geometric asymmetry in the detector which is
at the level of $\sim 10 \%$.}\label{fig2}
\end{figure}

Random daily weather-induced or instrumental variations of the anisotropy
itself (and therefore of FB) are averaged out by summing (i.e.\
averaging) many full days, each sweeping a full circle in the sky. The 
random anisotropies decrease as $\sqrt{N_{days}}$ while a coherent signal
is not reduced. Finally, because the method uses the time modulation of
FB rather than FB itself, it is impervious to the inherent geometrical
anisotropy present in the detector acceptance seen in the azimuthal 
distribution of Figure \ref{fig2}, which was observed to be stable 
during the lifetime of the experiment. We note that only instabilities
of this geometrical anisotropy on the scale of a single day would
affect the measurement, while a slow evolution would not.

Because this method measures the modulation in the direction of the Earth's rotation, it
yields no information about the modulation in the declination (dec.)
direction. The results will therefore be for the projection of the 
anisotropy in the right ascension (r.a.) direction rather than the full
2-D anisotropy of the sky. Such a projection can be created for
any visible dec.\ band. Since each
dec.\ band sweeps around a full circle of the sky independently
and contains statistically independent data, we choose to do separate
analyses for each $5^{\circ}$ dec.\ band of our data, considering each
as a separate observation. Any theoretical description of the true 2-D 
anisotropy can be confronted with, or constrained by, our data (given in
Section 4, Table 1) by projecting the anisotropy along r.a.\ in our 
dec.\ bands. In the rest of this paper we will always use the word anisotropy
to mean this projected anisotropy.

To do the harmonic fit we make the assumption that the large scale anisotropy
in any given dec.\ band can be modeled by a Fourier series and that it is a
small modulation of a nearly isotropic signal. Three harmonics
(the fundamental and the next two) have been found to be sufficient for this
method (see section 4.1). This allows us to see large scale effects having a
width in r.a.\ of greater than $\sim40^{\circ}$. In this harmonic model the
normalized rate in a given direction of the sky, in celestial equatorial 
coordinates with declination = $\delta$ and right ascension = $\theta$ (or
the equivalent coordinate in the right ascension direction when using
a different coordinate system) is:

\begin{equation}\label{skymod}
 \frac{R_{\delta}(\theta)}{<R_{\delta}(\theta)>} = 1 + A_{\delta}(\theta) = 1+\sum_{n=1}^3 \gamma_{n,\delta} \cos n(\theta - \phi_{n,\delta}) \\
\end{equation}

where for a given $\delta$ the average in the denominator of the left side is
over $\theta$, and $A_{\delta}(\theta)$ is the fractional anisotropy in the
right ascension direction. This model of the sky is complete once the
six Fourier coefficients (three amplitudes and three phases) on the 
right side of the equation are known. The data analysis thus consists
of determining these Fourier coefficients.

It is this harmonic model that allows us to connect the quantity of interest,
namely the fractional anisotropy in the sky $A_{\delta}(\theta)$, and our
chosen tool, the measurement of the modulation of the forward-backward
asymmetry. This connection is detailed explicitly below.

The quantity $N_{\theta_{0},\delta}(\xi)$ is the number of cosmic-ray events
collected during a particular time interval (characterized
by the angle $\theta_{0}$), in an angular bin at a given 
declination ($\delta$) and local hour angle ($\xi$) characterizing 
the symmetric forward ($+ \xi$) and backward ($- \xi$) inclination of the  
``telescope" of Fig.\ \ref{fig1}. In this notation the forward-backward asymmetry of Eq.
1 becomes:

\begin{equation}\label{eqn1}
  FB_{\delta}(\theta_{0}, \xi) = \frac{N_{\theta_{0},\delta}(+\xi) - N_{\theta_{0},\delta}(-\xi)}{N_{\theta_{0},\delta}(+\xi) + N_{\theta_{0},\delta}(-\xi)}
\end{equation}

The bin counts are computed for
$1/2$-hour intervals in sidereal time (ST), universal time (UT), and anti-sidereal time (AST) \citep{AST}. These $1/2$-hour intervals are parameterized by an angle
 $\theta_{0}$ which specifies the relative advance of the local meridian
through the sky for the three different time scales:

\[ \theta_{0} = 3.75^{\circ} + 7.5^{\circ}\times IST  \]
\[ \theta_{0} = 3.75^{\circ} + 7.5^{\circ}\times IUT  \]
\[ \theta_{0} = 3.75^{\circ} + 7.5^{\circ}\times IAST \]

IST, IUT, and IAST are integers, from zero
to 47, denoting half hour intervals of sidereal time, UT, and anti-sidereal
time (defined by flipping the sign of the transformation from universal 
time to sidereal time). In the above equations, the constants convert an
integration time interval (1/2 hour) into degrees ($7.5^{\circ}$) and a
starting angle at the center of that interval.

\begin{figure}
\epsscale{0.80}
\plotone{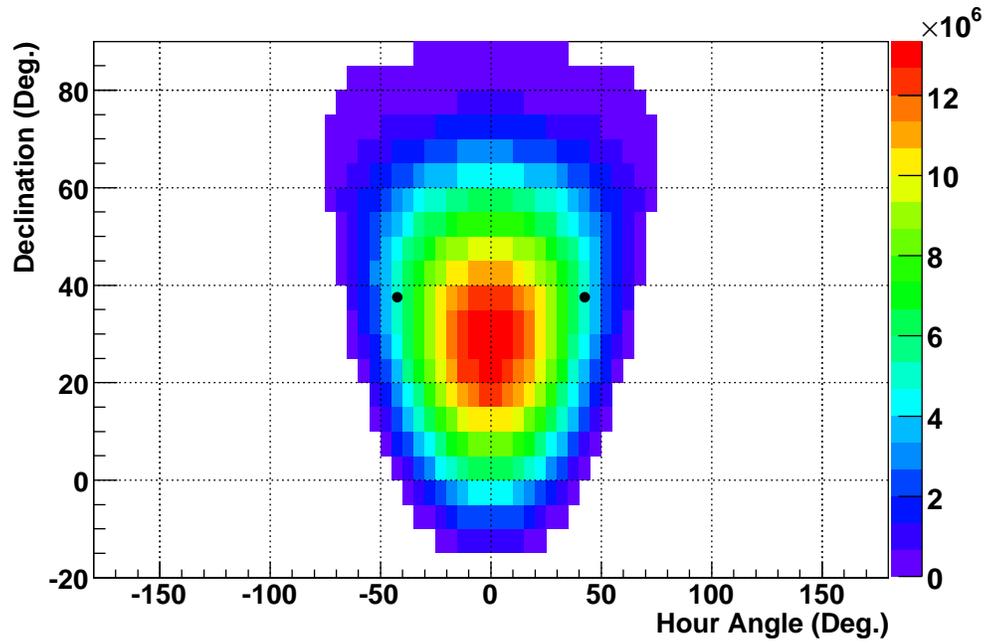}
\caption{Example histogram showing the number of cosmic-ray events vs.\ hour
angle and declination containing seven years of data for a single sidereal \
half-hour. The solid circles are an example pair of pixels shown in Fig. 1
(with $\xi = 42.5^{\circ}$) used in the calculation of the FB given by
Eq.(1) \& Eq.(3).}\label{fig3}
\end{figure}

In practice the cosmic-ray events are recorded in histograms with
$5^{\circ}\times5^{\circ}$ bins according to their arrival
direction from $-10^{\circ}$ to $80^{\circ}$ in declination and $-50^{\circ}$
to $+50^{\circ}$ in hour angle. The events are collected over a 30 ``minute"
period giving us 48 half hour histograms per day, where ``minute" is defined in the
three following time frames: sidereal 
(366.25 days/year), universal (365.25 days/year) and anti-sidereal
(364.25 days/year). The coordinate system in the sidereal time frame is
``sky-fixed", in the universal time frame it is 
``sun-fixed", and in the anti-sidereal frame it is ``non-physical". Histograms
for a given half hour period are accumulated over any number of days to build
an average set which corresponds to a chosen time period. A
representative example is shown in Figure \ref{fig3}. These 48 half hour
histograms for the time frame, period, and dec.\ band of interest are then
analyzed using the method of forward-backward asymmetry described above
combined with a harmonic fit.

Recognizing that $\theta = \theta_{0} \pm \xi$, substituting Eq. \ref{skymod}
for the harmonic sky model into Eq. \ref{eqn1}, and then utilizing $\gamma \ll 1$
and the appropriate trigonometric identities, we get:

\begin{equation}\label{fitfun}
FB_{\delta}(\theta_{0},\xi)\approx  \sum_{n=1}^3 - \gamma_{n, \delta} \sin (n\xi) \sin (n(\theta_{0}-\phi_{n, \delta}))\nonumber\end{equation}

Equation (\ref{fitfun}) is the prediction of the harmonic model of the sky, 
in terms of its Fourier coefficients, for the forward-backward asymmetry, to
be fit with our experimental data.

\begin{figure}
\centering

\begin{tabular}{ccc}

& \plotone{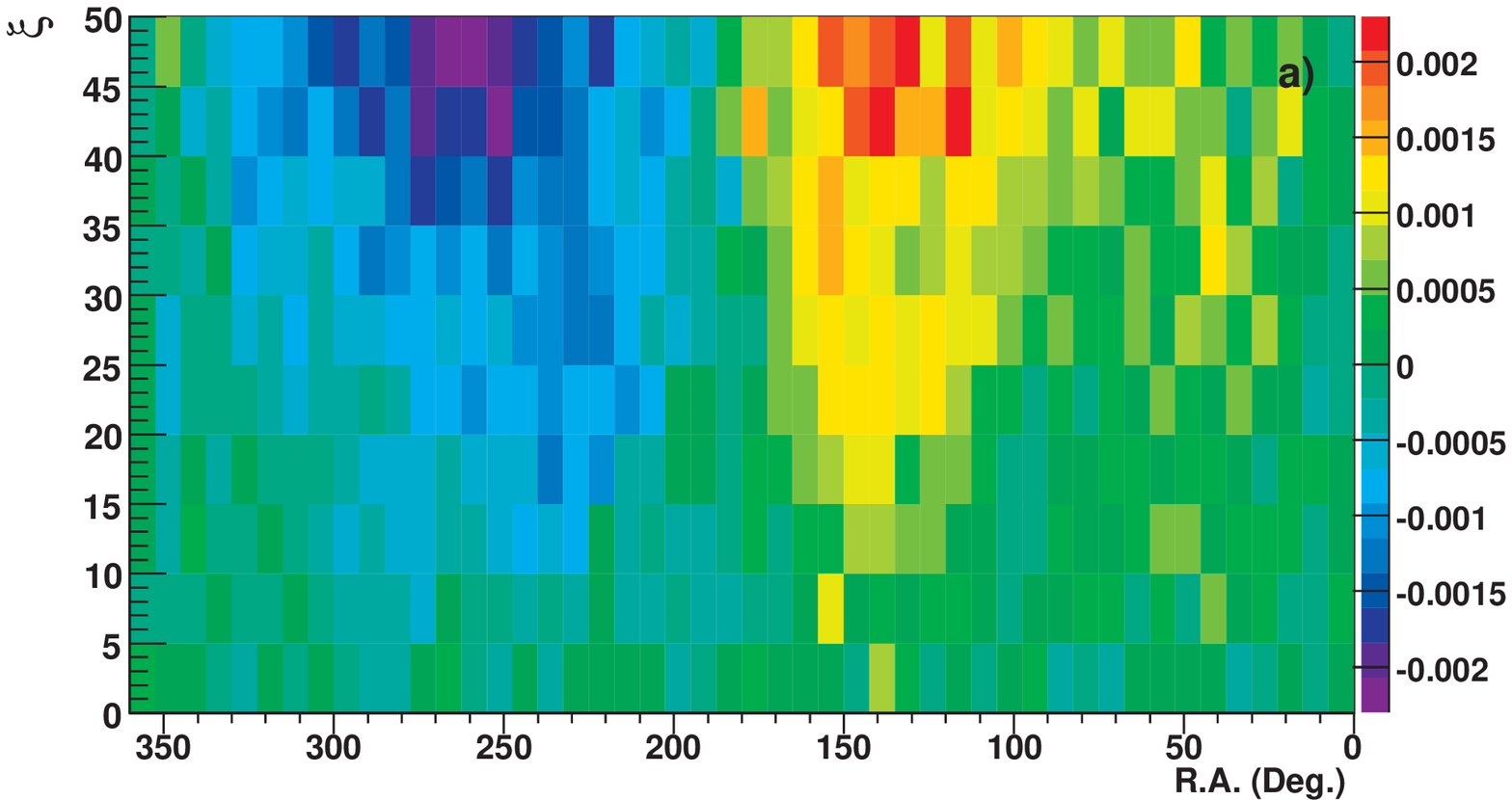} & \\
& \plotone{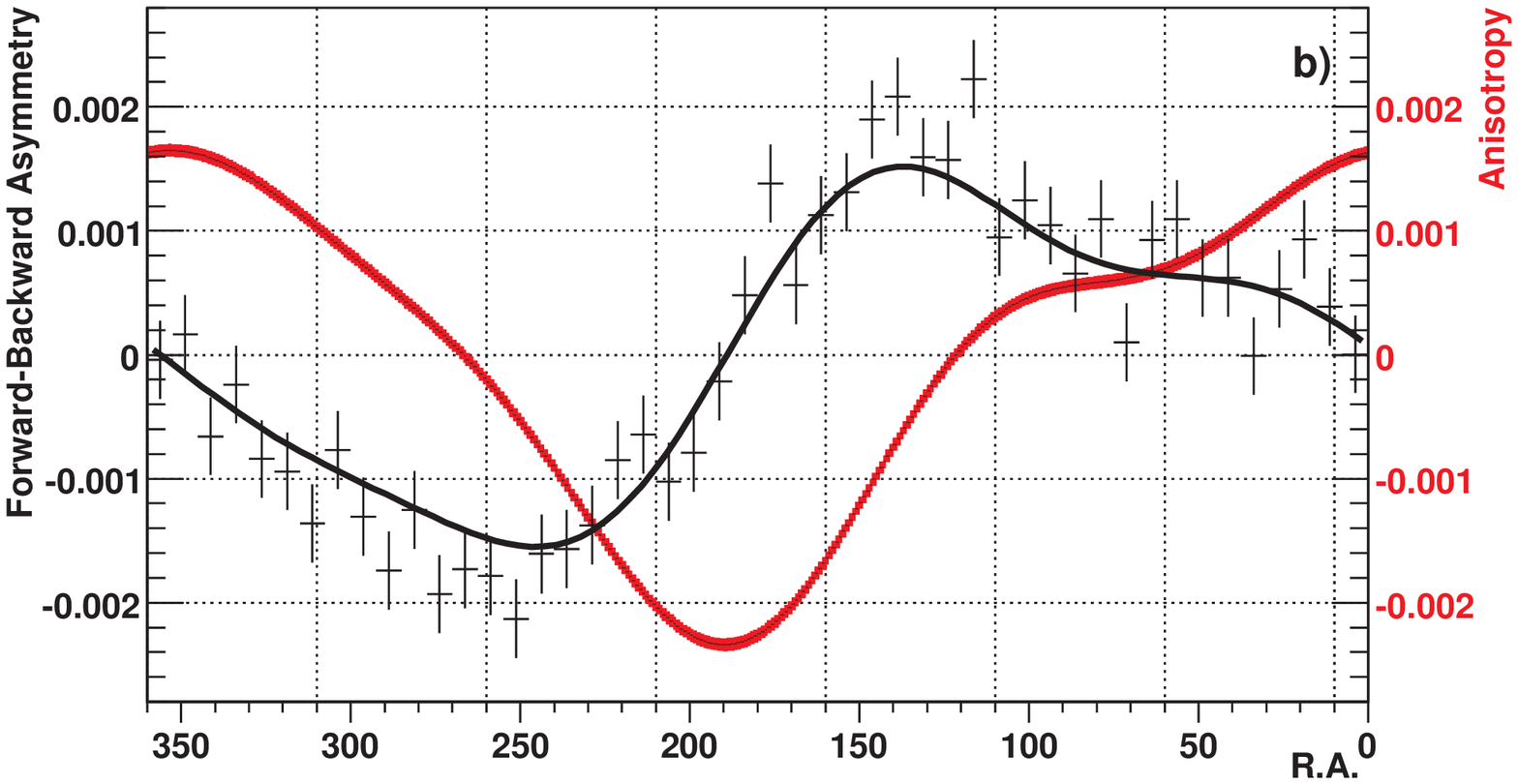} & \\

\end{tabular}


\caption{a) Sample 2-D histogram of FB vs. $\xi$ and r.a.\ for a single dec.\ 
band ($35^{\circ} - 40^{\circ}$) calculated using Eq.\ \ref{eqn1} for seven years 
worth of data (the binning reflects that the data is 
collected over half hour intervals which are parameterized by $\theta_{0}$). b) Sample histogram showing
the result of the 2-D fit of Eq.\ \ref{fitfun} to a) for a single slice 
in $\xi = 40^{\circ} - 45^{\circ}$ (in black, left axis), and the reconstructed 
anisotropy of Eq.\ \ref{skymod} using the six Fourier coefficients obtained by the 2-D fit to a) (in red, right axis).}\label{fig4}
\end{figure}

The next step is preparing the experimental data for the fit. For a fixed
``telescope" angle $\xi$ and at a given $\delta$ we calculate FB and
its statistical error by using the relevant pair of histogram bins in
each of our 48 saved histograms of hour angle vs. dec.\ (see Fig. \ref{fig3}
for an example). Since we are only interested in the modulation of FB with
$\theta_{0}$ (i.e.\ with time), the average over all $\theta_{0}$
is subtracted from the calculated FB. This step
separates FB due to the inherent geometric anisotropy of the detector
from the time modulated FB due to the sky anisotropy. These values are 
assembled into a 48 point, one-dimensional
histogram of FB and its error vs.\ $\theta_{0}$ for this particular
choice of $\xi$ and
$\delta$. Equation \ref{fitfun} could be fit to this histogram to
obtain the Fourier coefficients. But, this procedure can be done for
several choices of $\xi$ that we call $\xi_{i}$, each $\xi_{i}$ is
sampling the same sky, yielding a new FB vs.\ $\theta_{0}$ graph to fit. By
choosing a set of $\xi_{i}$ (where $\xi_{i}$ ranges from $2.5^{\circ}$ to a
dec.\ dependent maximum of up to $47.5^{\circ}$ in $5^{\circ}$ steps), centered on each of the
$5^{\circ}\times5^{\circ}$ pixels (see Fig. \ref{fig1} \& \ref{fig3}) and
calculating the FB (Eq.\ 3) with the contents of the corresponding pixels, we
obtain up to 480 (48 values of $\theta_{0}$ times up to 10 values of $\xi_{i}$) 
statistically independent measurements per dec.\ band. Every 
available pixel (histogram bin) pair in the 
48 hour angle vs.\ dec.\ histograms (Fig.\ \ref{fig3}) is used once and only
once. These one-dimensional histograms are assembled into a two-dimensional
histogram of FB vs.\ $\theta_{0}$ and $\xi_{i}$. This histogram contains
all our information about 
FB and its statistical error. The fit of Eq.\ \ref{fitfun} to this 2-D 
histogram (a simultaneous fit over $\theta_{0}$ and $\xi_{i}$) gives the 
experimental values and errors of the Fourier coefficients for the harmonic 
model of the anisotropy in a dec.\ band (Eq. \ref{skymod}). An example of the 
2-D histogram that is
being fit, and the result of the fit in $\theta_{0}$ for a single slice  
in $\xi_{i}$ are shown in Fig.\ \ref{fig4}(a) and (b). The coherent signal in
FB is clearly seen in both. Figure \ref{fig4}(b) also shows the resulting 
measured anisotropy reconstructed using the six fitted Fourier coefficients in
this example.

The determination of the statistical errors on the Fourier coefficients
is straight-forward. The statistical errors on the number 
of events ($N_{F}$ and $N_{B}$) are simply their square roots and these
errors are propagated to the error on FB:

\begin{equation}
\sigma_{FB} = \sqrt{\frac{1}{N_{F} + N_{B}} (1 + FB^{2})} \approx \sqrt{\frac{1}{N_{F} + N_{B}}}
\end{equation}

to a superb approximation. Since each FB bin in the 2-D histogram is 
statistically independent of the others, the fit is that of a 2-D experimental
histogram with independent errors to the parameters of the theoretical
function of Eq.\ (\ref{fitfun}). This fit (using MINUIT \citep{MIN}) propagates the 
errors on the experimental data into errors on the parameters of the theory,
namely the six Fourier coefficients of the anisotropy.

It is noted that the simultaneous fit of the different $\xi_{i}$ bands is
in effect an averaging over $\xi_{i}$ which means that we are also averaging 
over any difference of energy in the data due to the dependence of atmospheric
depth on $\xi$. This averaging is well justified after the fact by
an examination of the $\xi$ dependence discussed in Section \ref{stdataerr}.

Finally, we emphasize that the role of the forward-backward asymmetry
is over once the fit is done. The coefficients obtained are for the 
anisotropy of the sky, defined as the fractional difference of the rate
at a given r.a.\ from its average over all r.a., for each dec.\
band. In Section \ref{results}, we tabulate the fit results for these
18 dec.\ bands, as well as graphically display their behavior.

\subsection{Monte Carlo Tests of the Analysis Method}

Tests of the analysis method were conducted using a Monte Carlo simulation that 
takes as input a 2-D map containing any desired anisotropy in universal 
time (UT) or sidereal time (ST) and then outputs events which 
can then be analyzed using the method discussed in the previous section. 

\begin{figure}
\plotone{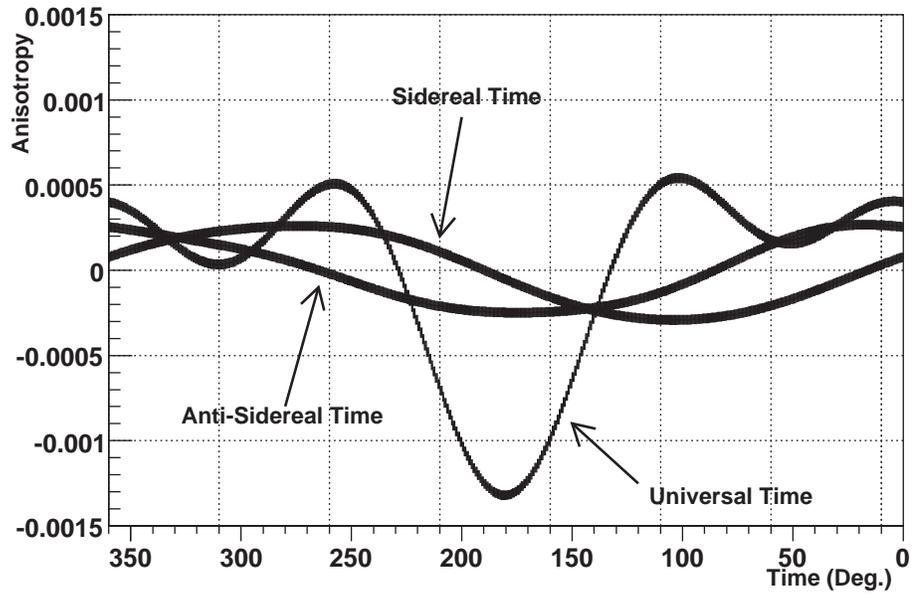}
\caption{Results of an all-dec-band anisotropy analysis using a MC
generated UT signal. The signal consists of square well deficit
from $150^{\circ}$ to $210^{\circ}$ in UT with an amplitude varying
sinusoidally in time from 0.000 to 0.003. The signal seen in the ST 
and AST frames is induced by the UT time variation. The width of the curves 
reflects the statistical error.}
\label{fig5}
\end{figure}

We describe one important case in detail: a fixed UT signal will average 
to zero in ST or AST, but a seasonally modulated day-night 
anisotropy in UT will generate a false ``sideband" signal in ST and 
AST. We use a time varying input in UT, where the magnitude of the UT 
signal varies sinusoidally between zero and its maximum with a period of one 
year, simulating an extreme seasonal variation. Figure \ref{fig5} shows the 
results of an all-dec-band analysis in three time frames for this 
test. The ST and AST frames, which have no input signal in the 
MC, show a clear signal induced by the time varying UT input. In this example 
the induced signal is at about half the amplitude of the UT time 
variation. The phases of the signals in ST and AST are different but their 
magnitudes are the same. Using this result we can estimate the systematic 
errors in the sidereal signal by 
examining the anti-sidereal signal. Since there are no physical processes 
which occur in the anti-sidereal time frame the signal should be zero when 
data sets of an integral number of years are used as seen from the Monte Carlo 
checks above. Therefore a signal present in the non-physical AST frame is 
interpreted as being induced by variations in time of the UT signal and such a 
signal will also appear in the ST frame. 

Two other sets of MC tests were successfully passed by the FB method. The 
first set tests the stability of the 
analysis method by using various simple signals (e.g.\ isotropic sky or the 
result of the analysis of real data) as input in ST with no input in UT or anti-sidereal 
time. Analysis of the output of these tests 
is consistent with the input within statistical errors. The UT and 
anti-sidereal time (AST) maps in these tests are consistent with isotropic 
as long as an integral number of years is used. This is to be expected 
since a static point in ST moves across the UT frame and returns to 
its original position after one year. In AST this point will move 
across two times in one year. Therefore when the declination band is 
normalized a sidereal signal will average to zero in both UT and AST. 
This cancellation is of course also true for the sidereal sky 
with a signal fixed in UT (e.g.\ day, night effects).

The second set of tests shows that no false harmonic features are 
induced from isolated structures (e.g.\ a square well deficit centered around 
$\sim185^{\circ}$ or the known Galactic equator excess of 
gamma-rays \citep{GalRidge1,GalRidge2,GalRidge3}). These tests show that no extra signals are 
induced except in the cases of a sharply discontinuous input and a time 
varying UT input. The discontinuous input induces features due to the use of 
only three harmonics as is expected from Fourier theory.

\section{Results}

\subsection{Multi-dec-band Results and the Sidereal Sky Map}
\label{results}

\begin{figure}
\plotone{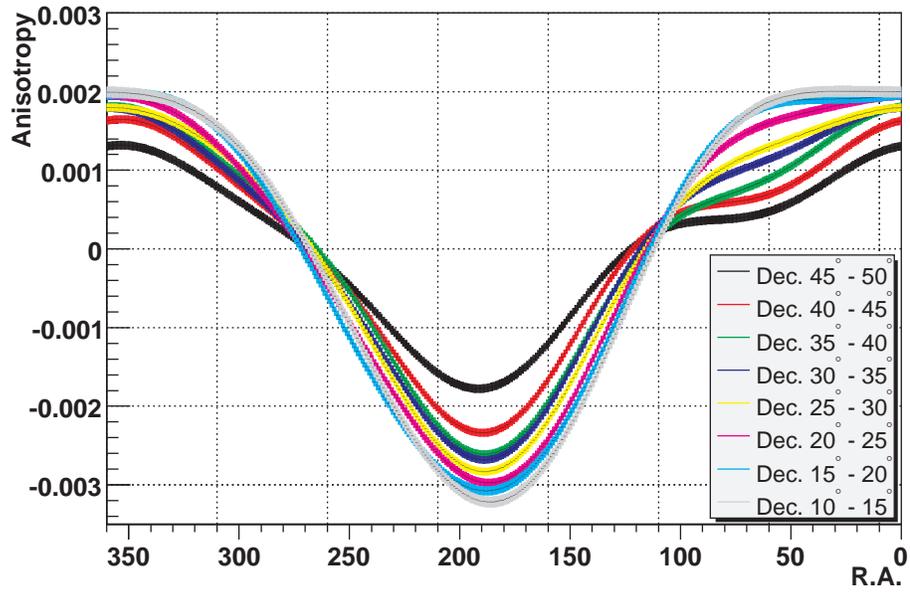}
\caption{Profiles in r.a.\ for individual $5^{\circ}$ dec.\ bands
from $10^{\circ}$ to $50^{\circ}$  used in the 2-D map seen in
Figure \ref{fig7}. The width of the lines reflects the statistical error.}
\label{fig6}
\end{figure}

  Table I is a summary of the harmonic fit parameters for the sidereal sky
fractional anisotropy obtained in each of 18 individual dec bands. The
$\chi^2$ for the fit and the sample size in each dec.\ band is also listed
in this table. Three harmonics were chosen as they give a better
$\chi^2$/ndf than one or two harmonics whereas there was little improvement
in using four. Fig. \ref{fig6} displays the anisotropy profiles in RA for 
eight adjacent dec.\ bands in the table. As seen from the definition of the 
anisotropy in Eq. \ref{skymod}, the profiles show the deviation from their 
average over all r.a., so that the area below and above the reference level 
of zero is the same: the existence of a 
deficit in some region implies an excess elsewhere.

\begin{figure}
\epsscale{0.80}
\plotone{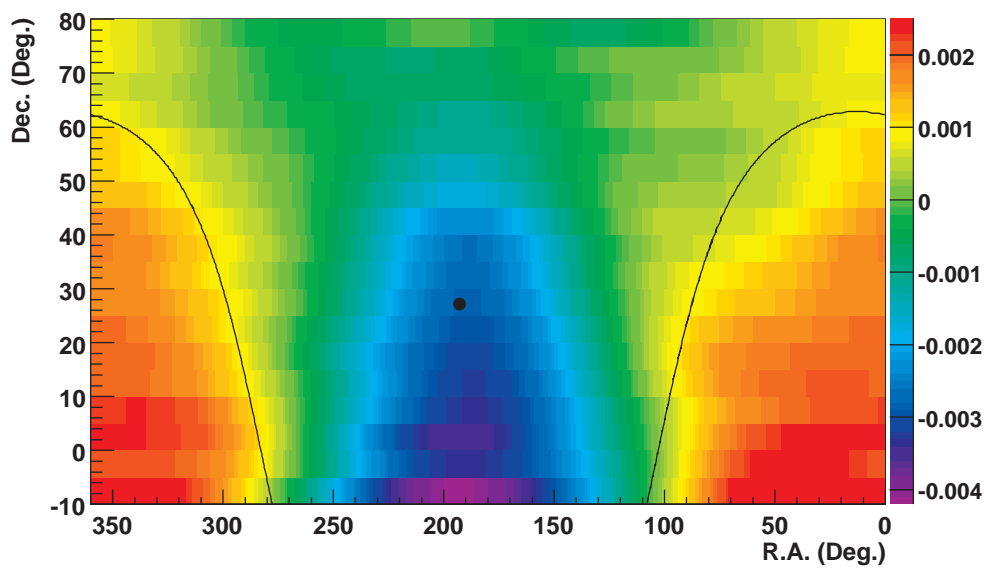}
\caption{Result of a harmonic fit to the fractional difference of the
cosmic-ray rates from isotropic in equatorial coordinates as viewed by
Milagro for the years 2000-2007. The color bin width is $1.0\times10^{-4}$
reflecting the average statistical error. The two black lines show the position
of the Galactic Equator and the solid circle shows the position of the
Galactic North Pole. This map is constructed by combining 18 individual
profiles of the anisotropy projection in r.a.\ of width $5^{\circ}$ in
dec. It is not a 2-dimensional map of the sky. The median energy of the events in this map is 6 TeV.}
\label{fig7}
\end{figure}

Figure \ref{fig7} assembles all 18 anisotropy profiles in their
respective dec.\ positions into a 2-dimensional color display of the local
anisotropy, with red representing an excess and blue representing a 
deficit. Care must be taken in interpreting Fig.\ \ref{fig7}. As 
explained before, our measurement is in the r.a.\ direction only, so that this 
picture does not purport to be the complete anisotropy of the celestial 
sphere; nor can the complete anisotropy be inferred from our 
results. Fig.\ \ref{fig7} is a two-dimensional display of one-dimensional 
information, the fractional
anisotropy in r.a.  What can be fairly compared across the dec bands
are the shape and strength of the fractional anisotropy in the r.a.\
direction; there is no information on the
r.a.\ averaged cosmic ray rate difference from one dec.\ band to
another, i.e.\ the anisotropy in the dec.\ direction.

\begin{figure}
\plotone{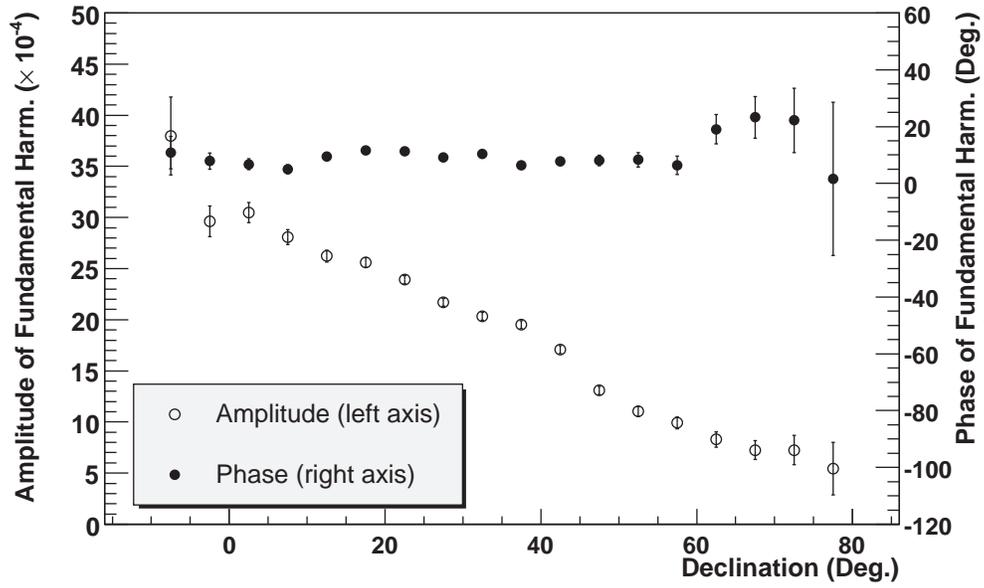}
\caption{Amplitude and phase of the fundamental harmonic obtained from a fit
to seven years of data for each $5^{\circ}$ declination slice. The error bars
are statistical.}\label{fig8}
\end{figure}

The data and their analysis are completely independent for each dec.\
band, but a striking commonality of the anisotropy behavior is seen in
both Figs. \ref{fig6} and \ref{fig7}. Adjacent dec.\ bands have very
similar profiles, in shape and in strength. The conclusion is that there is
indeed a coherent anisotropy signal over a large swath of the sky. The
dominant feature is a prominent valley or deficit region extending from
$150^{\circ}$ to $225^{\circ}$ in r.a., and clearly visible in all dec.\ bands
between $-10^{\circ}$ and $45^{\circ}$. The decrease of the depth of this
valley, towards large values of dec., seen in the color picture of 
Fig.\ \ref{fig7}, is explained at least in part by the fact that the 
circle in the celestial sphere generated by the earth's rotation gets smaller 
and smaller, shrinking to a single point as dec.\ approaches $90^{\circ}$. The 
r.a.\ position of the valley minimum appears dec.\ independent in 
Figs.\ \ref{fig6} and \ref{fig7}. Figure \ref{fig8} plots the 
amplitude and phase of the lowest harmonic versus dec., confirming both 
these trends; the amplitude is seen to decrease as dec.\ gets larger exhibiting 
the expected cos(dec) dependence, while 
the phase is quite stable for dec.\ below $60^{\circ}$, where the 
amplitude is large enough for the phase to be well defined.

\subsection{Single-dec-band Results for ST, UT, and AST Time Frames}
\label{sbresults}

\begin{figure}
\plotone{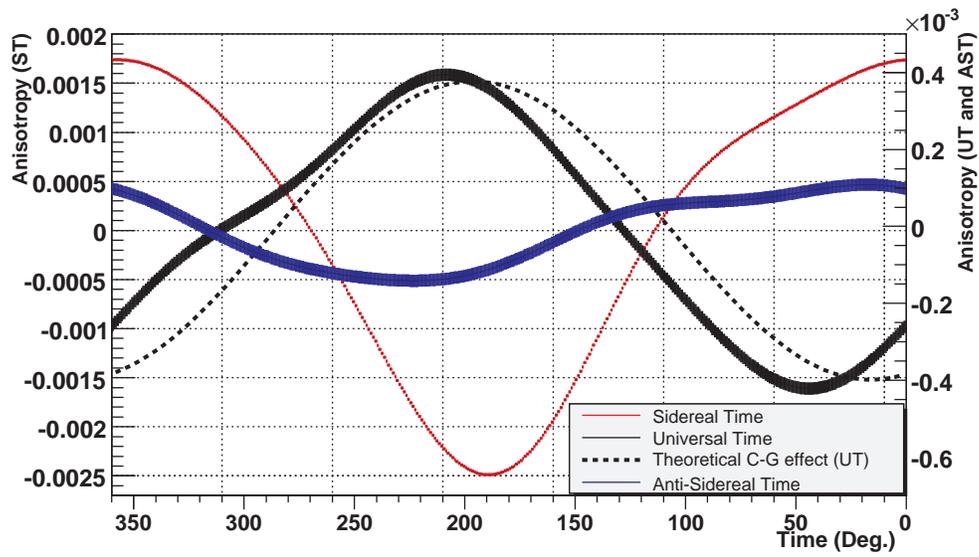}
\caption{Anisotropy constructed by fitting a single projection containing
data collected from all declinations collected over seven years in ST (red, 
left axis), UT (black, right axis) and AST (blue, right axis). The right axis 
(UT \& AST) is expanded by a factor of 4 compared to the left axis (ST). The 
dashed curve is 
the predicted Compton-Getting (C-G) effect due to the
Earth's motion around the sun which has an expected maximum value at 6 hr local solar 
time ($196.3^{\circ}$ UT given the longitude of Milagro). In these plots the 
width of the curve reflects the statistical error.}\label{fig9}
\end{figure}

As a complement to the multi-dec-band analysis of section 4.1
we do a single-band analysis by projecting the hour angle vs.\ dec.\ histograms onto
the horizontal axis of Fig.\ \ref{fig3}, thus integrating all dec.\ bands, then
performing the same anisotropy analysis on this single band, to
obtain the overall r.a.\ behavior. Examining the resulting single band
anisotropy, with only 6 fourier coefficients, gives a convenient and
economical way to compare and contrast the behavior in all three coordinate
systems corresponding to sidereal, universal and anti-sidereal time. Table 2
gives the fitted harmonic fit coefficients for ST, UT, and AST, and
Fig.\ \ref{fig9} displays the resulting profiles for all three
coordinate systems.

The sidereal-frame profile of Fig.\ \ref{fig9} clearly reproduces the
dominant feature seen in the multiband analysis, the valley centered at
$189^{\circ}$ in r.a. The single band valley depth (SBVD) is an alternate
measure of the strength of the sidereal anisotropy and is found to be SBVD =
(2.49 $\pm$ 0.02 stat.\ $\pm$ 0.09 sys.) $\times 10^{-3}$, a
$22.6\sigma$ signal. The method of systematic error estimation will be discussed 
in Section \ref{stdataerr}.
 
The Compton Getting effect due to the earth's motion around the sun gives
a predicted anisotropy in universal-time frame which is a dipole
(the lowest harmonic), plotted in Fig. \ref{fig9} together with
our experimental result. Because the predicted signal amplitude is a factor 
of five smaller than the observed sidereal time signal, and is 
correspondingly more susceptible to systematic distortions, this is a 
stringent test of the FB method. The experimental anisotropy profile is 
indeed dominated by the lowest harmonic as expected, has the predicted 
amplitude and reasonably close, though not in agreement (within errors) with 
the predicted phase. A detailed analysis of 
the UT measurement, including systematic errors, is deferred 
to a future publication. We consider our observation of 
the Earth-motion Compton-Getting anisotropy signal in UT a powerful check 
of the analysis method, and thus a verification of the reliability of the 
result for the sidereal sky.

Since the anti-sidereal reference frame of Fig. \ref{fig9} is
non-physical, we expect no real signal there, while systematic effect
distortions will show up in this frame also. Indeed this signal, plotted in Fig.\ \ref{fig9}, is seen to be a factor
of 3 smaller than the UT signal and thus a factor of 15 smaller than the
ST signal. The observed AST signal will be used in the estimation of systematic
errors in the next section.

\subsection{Stability of Data and Systematic Errors}
\label{stdataerr}

\begin{figure}

\plotone{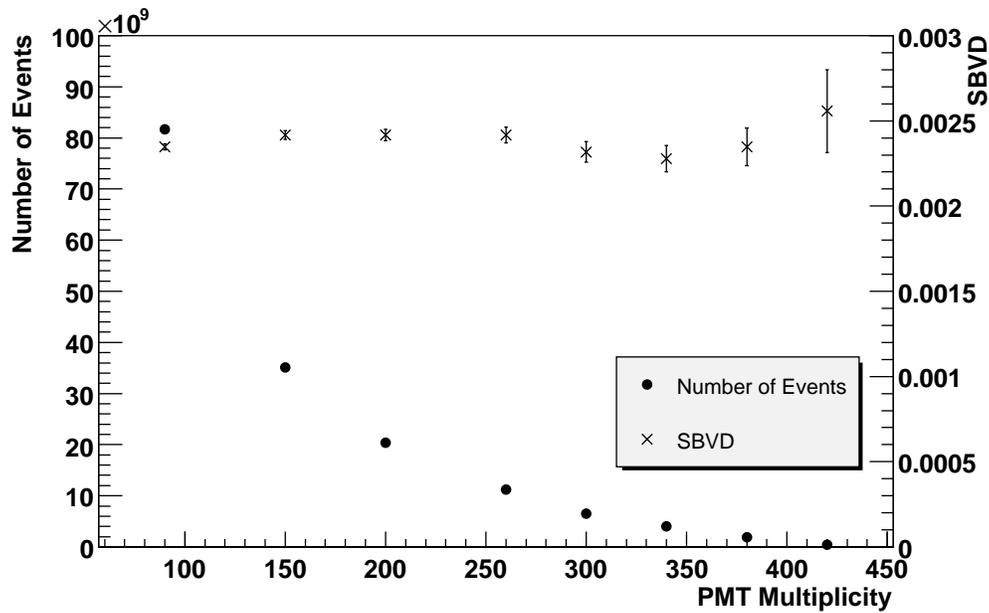}

\caption{Total number of events over a six year period (left axis) and valley 
depth SBVD (right axis) vs.\ PMT multiplicity. The PMT multiplicity is defined as being the number of PMTs hit 
in the top layer of the pond for a given event. The lack of change in SBVD 
seen compared to the large decrease in the number of triggers shows explicitly 
the insensitivity of the analysis to variations in the trigger threshold.} 
\label{fig10}
\end{figure}

As a verification of the threshold independence and overall detector rate 
independence of the method and results, we raise the PMT multiplicity 
(the number of PMTs hit during an event in the air shower layer of the 
detector) requirement in the trigger from its hardware value of around 75 
in several steps up to a value of about 400. Figure \ref{fig10} shows 
the number of events and the fractional 
anisotropy valley depth SBVD as a function of the PMT multiplicity. The valley 
depth is seen to be stable over a factor of five in trigger threshold 
corresponding to a factor of $\sim 25$ in trigger rate. This is a direct 
confirmation that the result is indeed impervious to the detector rate and 
threshold, as expected from the rate cancellation of the FB method, explained 
in Section \ref{fbahm}. 

\begin{figure}
\plotone{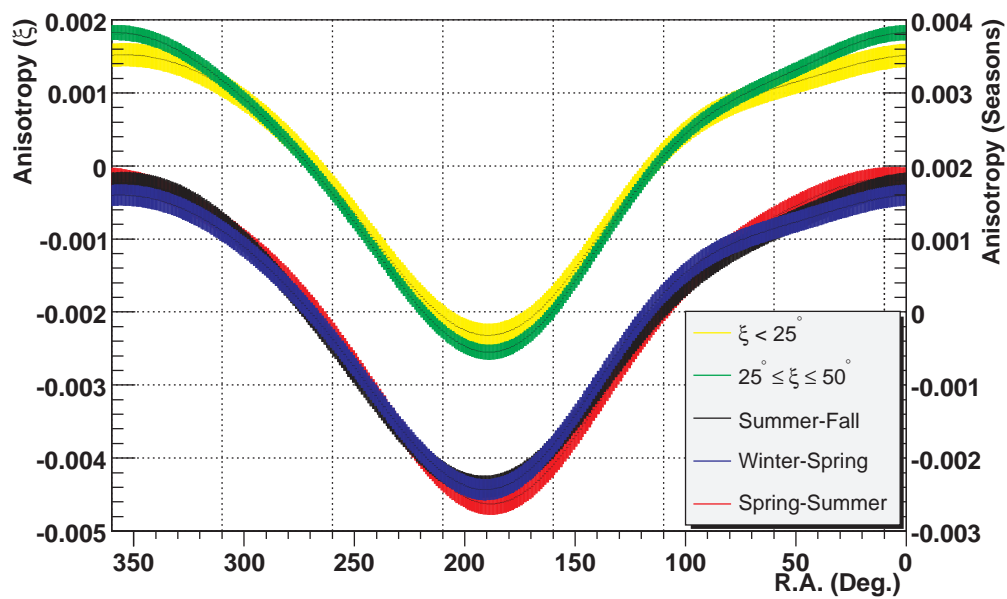}
\caption{Single-dec-band analysis showing seven years of data
split into two sets: $\xi < 25^{\circ}$ (in yellow), and $25^{\circ} \leq \xi
\leq 50^{\circ}$ (in green). Also plotted is the seven years of
data split into three seasonal sets: Summer-Fall in black, Winter-Spring in
blue, and Spring-Summer in red. The error bars are statistical + 
systematic. The independence of the analysis on $\xi$ as well as 
the lack of seasonal variation can be seen here. }
\label{fig11}
\end{figure}

In section 3.1 it is mentioned that the analysis method averages over different values of 
$\xi$, from $2.5^{\circ}$ to $47.5^{\circ}$. Figure \ref{fig11} compares 
the single-band profiles when the full data set is split in two and 
analyzed separately, where one set contains data with $\xi$ ranging 
from $0^{\circ}$ to $25^{\circ}$ and the other $25^{\circ}$ 
to $50^{\circ}$. The good agreement of these two profiles demonstrates that 
the procedure of simultaneously fitting over all $\xi$ is well justified.

To check for the possibility of a dependence on seasons, we break the data 
into three seasonal sets, defined as: Apr.-Jul., Jul.-Nov., and 
Nov.-Apr.\ corresponding to average local weather periods of: warm with low 
precipitation, high precipitation, and ice and snow respectively. No 
significant changes are seen in Fig.\ \ref{fig11} which shows the three 
single-band profiles superimposed.

\begin{figure}
\epsscale{0.80}
\plotone{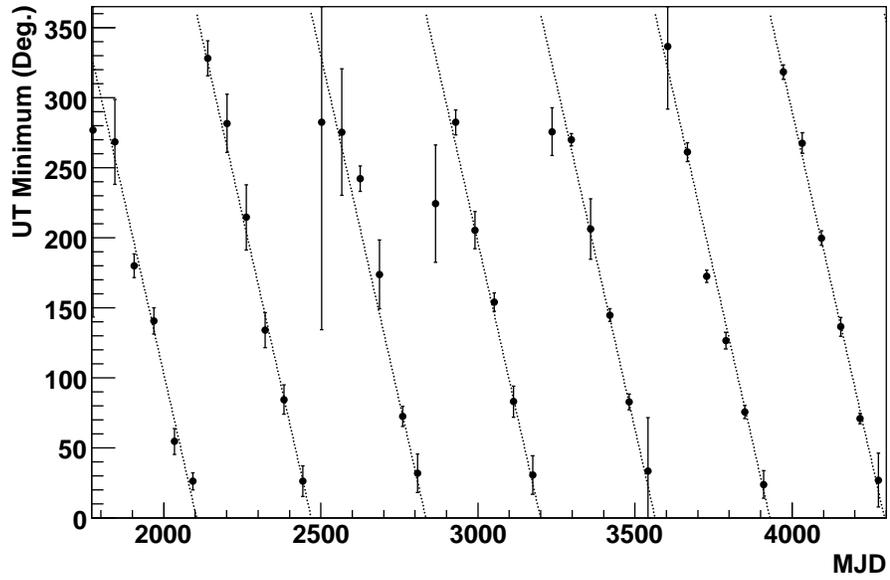}
\caption{Position of minima in universal time vs.\ Modified Julian Date (MJD) 
for data collected over the 42 two-month periods starting in July 2000 and ending in July 2007. The error bars are the stat.\ errors. The solid line is the 
calculated position in universal time of a constant sidereal position at r.a.\ 
equal to $189^{\circ}$. The excellent agreement between the data and the 
calculation shows that the UT signal is dominated by the ST signal.}
\label{fig12}
\end{figure}

To test whether the observed ST signal is a sidereal phenomena
and not due to universal time effects, such as weather, we break the data into two
month periods. On these short time scales there will not be a cancellation
between the ST, UT and AST time frames since a fixed position in ST
corresponds to a nearly constant position in UT. Fig.\ \ref{fig12} shows the position of
the observed minima in UT for each two month period over seven years. The solid
line shows the expected position of a fixed sidereal minimum located at $189^{\circ}$ r.a. If
the minimum observed at $189^{\circ}$ r.a. is a dominant feature of the
sky, then the minimum in UT should follow the sawtooth pattern
of the solid lines in the figure. If, on the other hand, there were dominant
features in the UT sky, the positions of the minima should not correlate at all
with this sawtooth pattern. The observed correlation between the observed and
predicted positions of the minima in Figure \ref{fig12} is a strong indication that
the ST signal is the dominant feature in UT on this time scale. The same
conclusion is reached when the corresponding analysis is carried out in the
AST time frame where the sawtooth pattern has double the frequency.
 
\begin{figure}
\epsscale{0.80}
\plotone{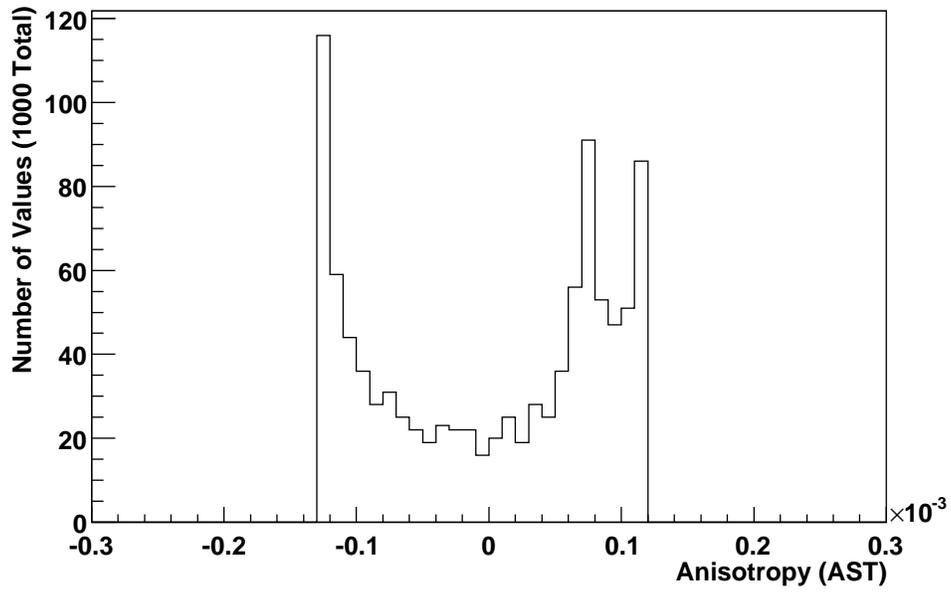} 
\caption{Histogram showing the projection of the AST curve from 
Figure \ref{fig9} onto the anisotropy axis. The RMS of this distribution 
is $0.09 \times 10^{-3} $ which is used to determine the systematic error on 
the sidereal signal.}\label{fig13}
\end{figure}

 To estimate systematic errors in ST or UT, for data sets
of a full year (or several full years) of data, we use the
anisotropy profiles in AST. The phase in the nonphysical
anti-sidereal frame sweeps continuously between $0^{\circ}$ and $360^{\circ}$
in ST or UT in the course of a year, so that any coherent signal is lost
because its phase is in effect randomized by this sweep. Since the
physical signals have been washed out, what is left are any systematic
distortions. We illustrate the logic and procedure
for the case of the AST profile seen in Fig.\ \ref{fig9} for the single band
analysis.  While the deviation of this profile from zero gives the possible 
magnitudes of the distortion of the valley depth, the phase relative 
to the signal in ST is unknown. If a deficit region of distortion in the 
AST profile were to coincide with the valley
it would increase the valley depth SBVD, an excess would decrease
it, and a zero crossing would leave it unchanged. We account
for all these possibilities by projecting the AST curve of Fig.\ \ref{fig9} onto the vertical axis, to obtain the histogram shown in Fig.\ \ref{fig13} of all the
possible distortions of SBVD that can result from this profile. We then
take the RMS deviation of this histogram, $\sigma_{RMS} = 0.09 \times 10^{-3}$ 
as the systematic error on the valley depth SBVD in the ST analysis.

\subsection{Time Evolution of the Sidereal Anisotropy}
\label{timeev}

\begin{figure}
\centering
\begin{tabular}{ccc}

& \plotone{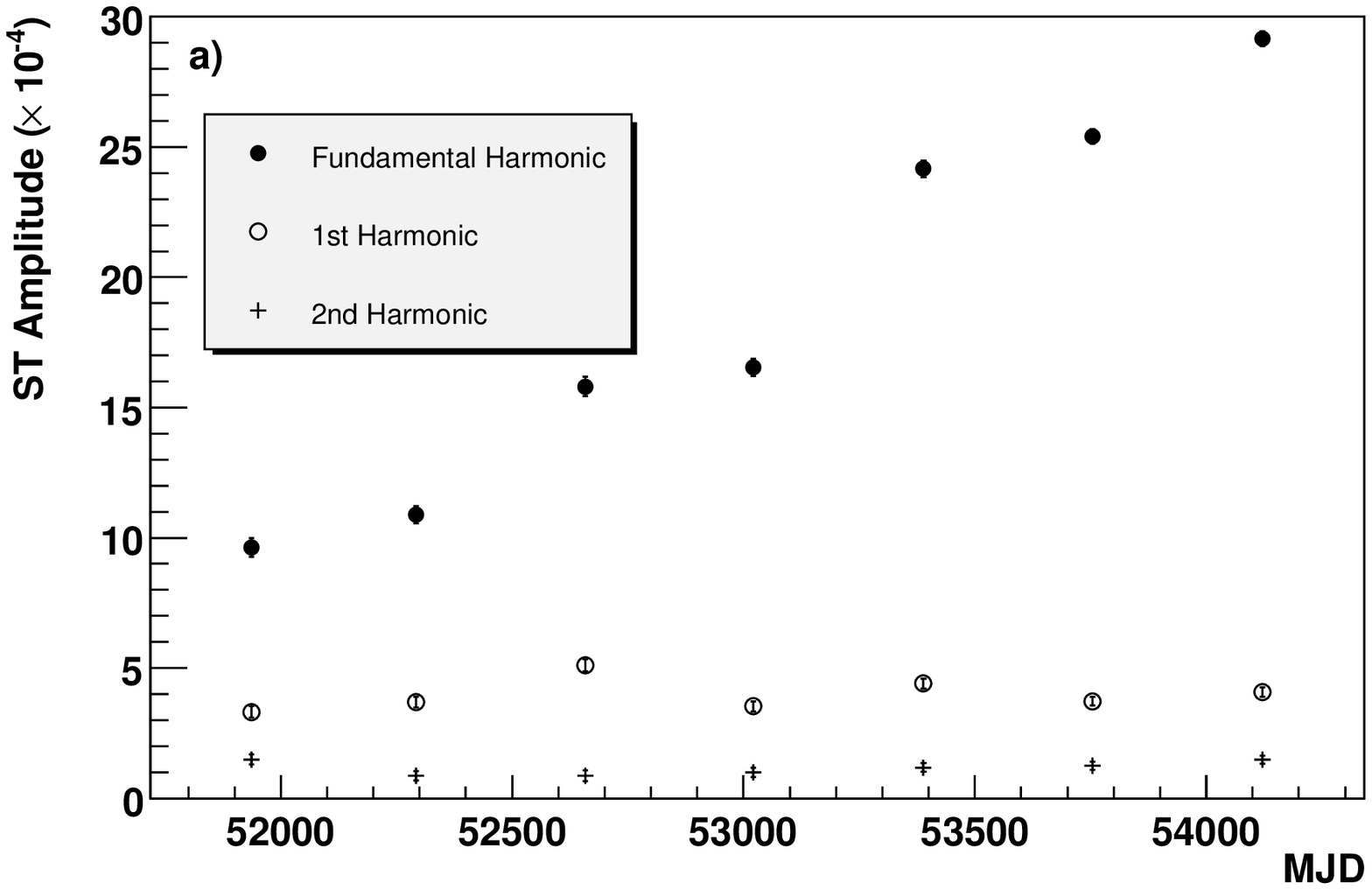} & \\
& \plotone{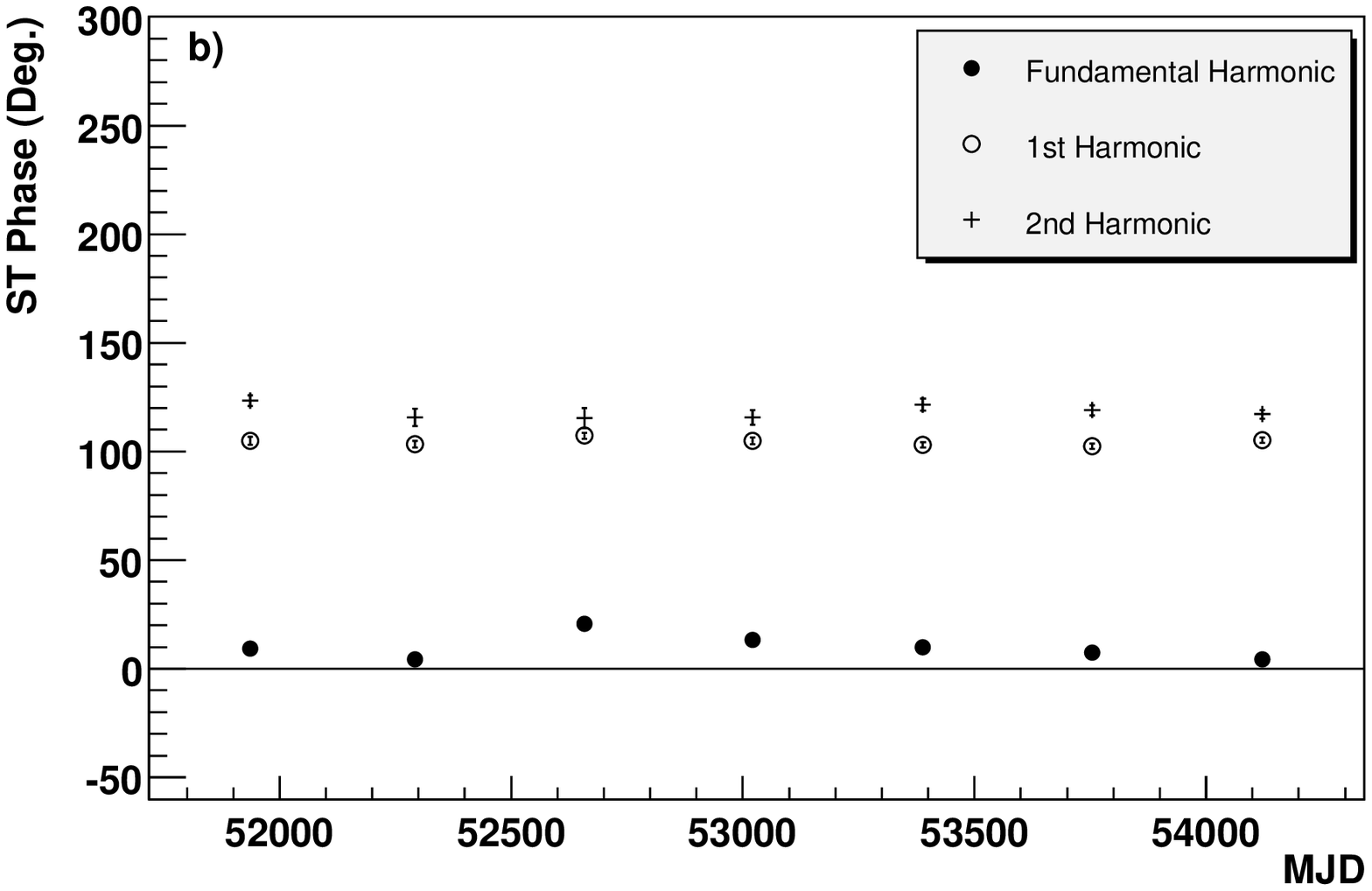} & \\

\end{tabular}
\caption{a) Sidereal time amplitudes of the three fit harmonics for the single band (all dec.) analysis. b) Sidereal time phases of the three fit harmonics for the single band analysis. Both plots contain seven yearly data sets from July 2000 to July 2007. The 
error bars are statistical.}
\label{fig14}
\end{figure}

To examine any time evolution of the anisotropy the data were split into
seven year-long sets from July 2000 to July 2007. Figure \ref{fig14} shows the
three amplitudes and three phases obtained by fitting data from each yearly
set using the single band analysis (over all dec.). There is a stability
in phase of all three harmonics over these seven years and in amplitude of the
two higher harmonics. The fundamental harmonic, however, shows a clear increase in
amplitude with time. 

In order to quantify this increase in a
precise manner, we look at the time evolution of the single band valley depth
(SBVD) in the all dec.\ analysis, defined in
Sec.\ \ref{sbresults} as a single-parameter measure of the strength of
the anisotropy. Using this parameter allows us to include all three
harmonics as well as utilize the systematic error estimation procedure
outlined in the previous section. First we confirm that the valley 
is stable in position over time. The position in r.a.\ of the minima 
over the course of seven years shows very little variation as is expected 
from the stability of the harmonic phases seen in Fig.\ \ref{fig14}. Fitting 
the position of the seven yearly minima to a constant yields a value 
of $189^{\circ} \pm 1^{\circ}$ r.a.\ with a $\chi^{2}/ndf = 4.5/6$. This 
lack of change in position over time is what one would expect from an actual 
sidereal signal.

\begin{figure}
\epsscale{0.80}
\plotone{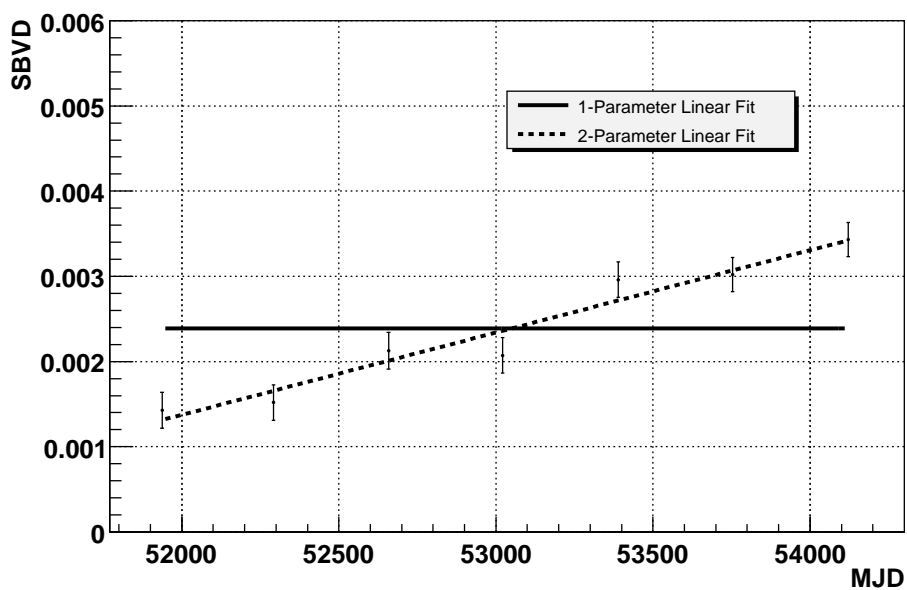}
\caption{Valley depth in the all-dec-band analysis 
(SBVD) vs.\ MJD for yearly sets from July 2000 to July 2007. The error bars are the linear 
sum of the statistical \& systematic errors. The solid line is the fit
to a constant value and the dashed is the linear two-parameter fit. The
$\chi^{2}$/ndf for the fits are 86.2/6 and 4.4/5 respectively. The fit
parameter in the flat case is $(2.39\pm0.08)\times10^{-3}$; the two fit 
parameters to the function $A(MJD)=p_0(MJD-53000)+p_1$ are: 
$p_0 = (0.97\pm0.11)\times10^{-6}$ and $p_1 = (2.34\pm0.08)\times10^{-3}$.}
\label{fig15}
\end{figure}

Figure \ref{fig15}, SBVD vs.\ year, shows that there is strong
evidence of a strengthening of the valley depth over the seven year
span of this data set. 

To test the robustness of this time dependence a number of checks were
done. As a test that is completely different from the insensitivity of 
anisotropy strength to trigger 
thresholds (described in Section \ref{stdataerr} and Fig.\ \ref{fig10}), we 
have done a direct check of whether the time-dependence of SBVD itself is 
threshold dependent. For several raised multiplicity thresholds between 90 
and 280 PMTs hit in the top layer of the pond, the same time dependence is seen; the 
yearly trend does not disappear. 

\begin{figure}
\centering
\begin{tabular}{ccc}

& \plotone{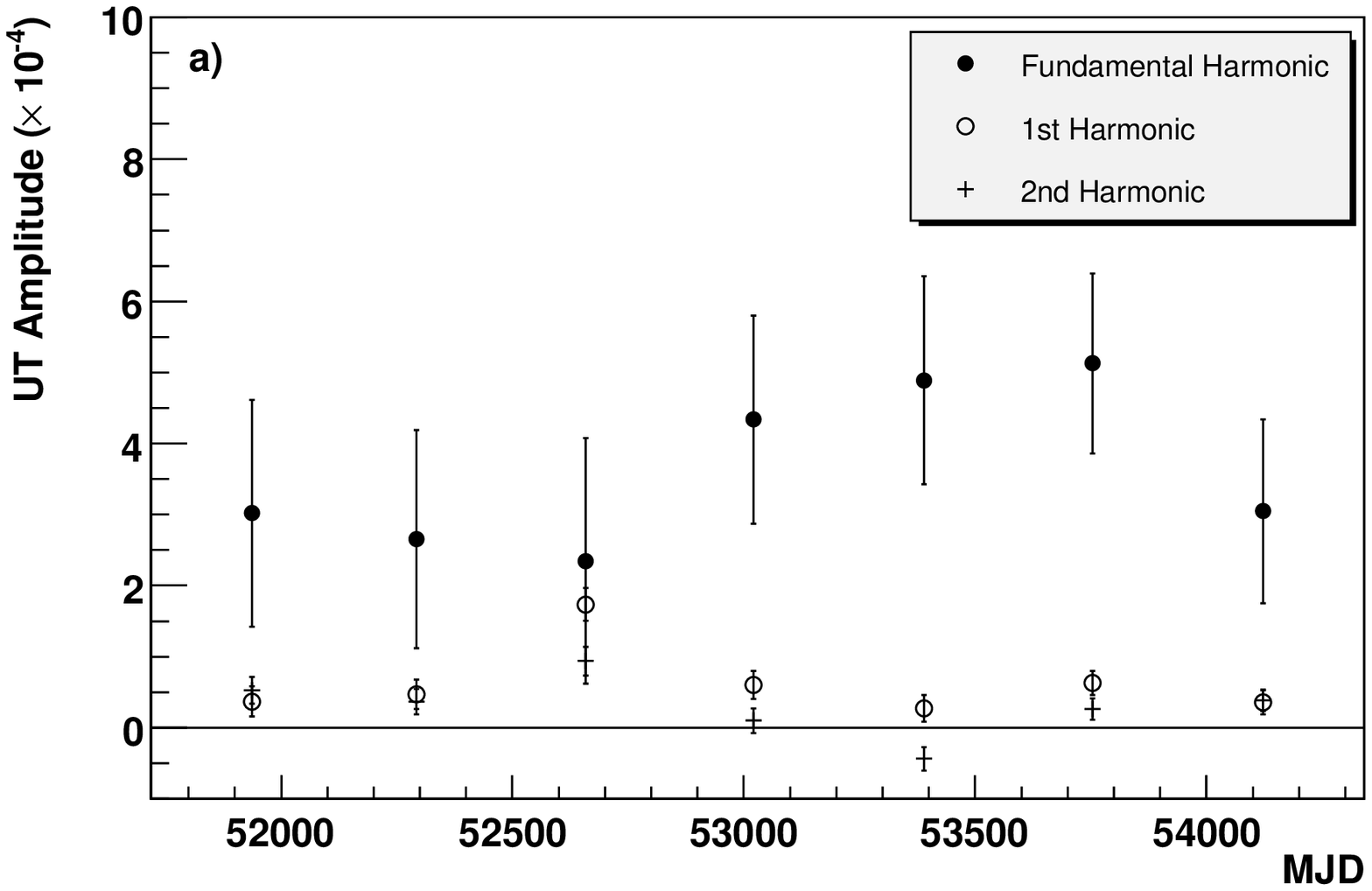} & \\
& \plotone{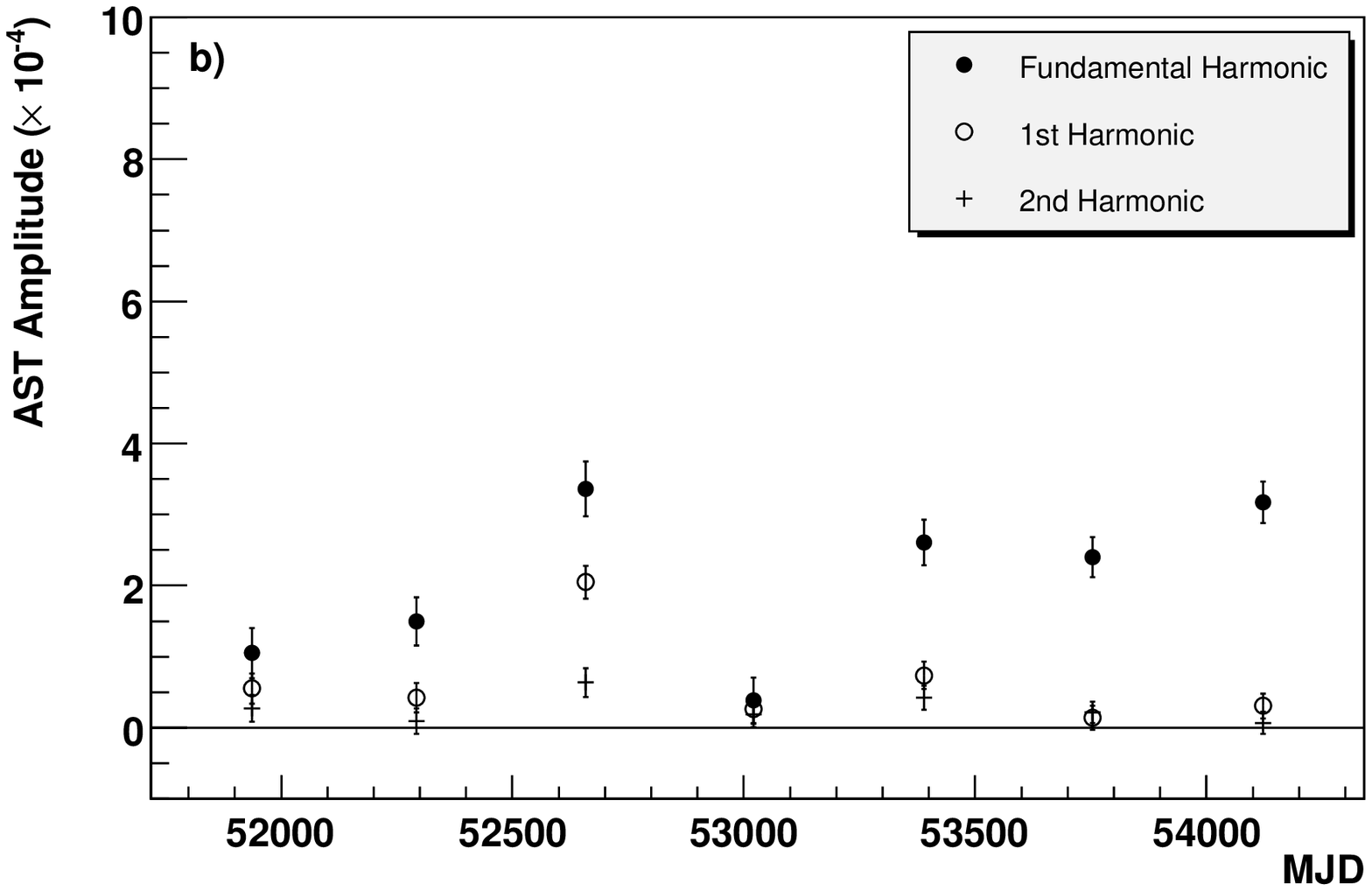} & \\

\end{tabular}
\caption{a) Universal time fit amplitudes for the single band (all dec.) analysis for seven yearly data sets from July 2000 to July 2007. b) Anti-sidereal time fit amplitudes for the single
band analysis for yearly data sets. For the UT fundamental harmonic only we 
show the statistical error + an estimate of the systematic error. For AST the 
error bars are only statistical. Note the 
lack of any definite trend, as opposed to what is seen in ST (Fig. \ref{fig14})}
\label{fig16}
\end{figure}

To see that this is a sidereal effect and not a detector
effect we look at the yearly time evolution for the universal time and
anti-sidereal time signals. Figure \ref{fig16} shows the amplitudes of the
three fit parameters for the single band analysis (all dec.) in both UT and
AST; the earth motion Compton-Getting effect in UT should have no
time dependence of the amplitude. The amplitudes of the harmonics in UT 
are constant over this seven year data set, within the errors, as well 
as their phases (not shown in the figure). With respect to the amplitudes of 
the harmonics in AST, these appear to be significantly larger in some 
years, but even the largest amplitudes are 5 to 10 times less than those 
seen in ST. From these tests it thus appears that time dependent
detector effects cannot account for the observed strong time dependence of the
sidereal anisotropy.

\section{Conclusions}

Previous experiments such as the Tibet Air Shower Array, with a modal energy of 
3 TeV, and Super-Kamiokande-I, with a median energy of 10 TeV,  have identified two coincident regions of
interest in their sidereal observations: an excess located at
 $\sim75^{\circ}$ r.a.\ or ``tail-in" anisotropy, and a deficit at 
$\sim200^{\circ}$ r.a.\ or ``loss-cone" anisotropy \citep{TB,SK}. 
Both of these regions are consistent with Milagro observations.
The ``loss-cone" coincides with the deep central-deficit region seen in 
this analysis while the narrow ``tail-in" excess is clearly observed in another Milagro analysis which is 
sensitive to features with extent smaller than $\sim30^{\circ}$ in r.a. (see \citet{GW} region A). 

The strengthening of the signal in the central-deficit region over time 
is a result unique to this analysis. Tibet found no evidence of time variation 
comparing data split into two five-year periods, 1997-2001 and 
2001-2005. However, only the second of these time periods overlaps with our data set for which 
the average value we observe in the deficit region agrees 
well with their measured deficit.

The anisotropy observed in the galactic cosmic rays could arise from a number of 
possible effects. The Compton-Getting effect (CG) predicts that due to the 
motion of the solar system around the galactic center through the rest frame 
of the cosmic-ray plasma an anisotropy is induced with the form of a dipole 
with a maximum in the direction of motion. For no co-rotation of the cosmic 
ray plasma with the Galaxy, the magnitude 
of the anisotropy is calculated to be $0.35\%$ given our speed of 
$\sim220 km s^{-1}$, while at the other extreme of full co-rotation, the 
anisotropy would be zero. No evidence of a Galactic CG anisotropy was seen in 
\cite{TB}. For no co-rotation, the dipole should have a maximum at r.a.\ = $315^{\circ}$ 
and dec.\ = $48^{\circ}$, and a minimum at r.a.\ = $135^{\circ}$ and 
dec.\ = $-48^{\circ}$. Since our analysis method yields only a projection of the anisotropy the observed CG effect will be slightly different from the true 
effect. The CG effect we expect to see 
in this analysis was determined from Monte Carlo simulation and found to be 
a dipole with a maximum of $0.14\%$ at $315^{\circ}$ r.a.\ for the declination 
range between $50^{\circ}$ to $60^{\circ}$. This range was considered to try 
to reduce effects from the central-deficit region. For the actual data, 
fitting a single harmonic to the projection in r.a.\ corresponding 
to declinations $50^{\circ}$ to $60^{\circ}$ gives a $\chi^{2}/d.o.f.\ = 11505/998$ 
which is clearly a poor fit. Although this suggests that the observed 
anisotropy is not dominated by the galactic 
Compton-Getting effect, its contribution to the anisotropy cannot be 
ruled out.

In addition to the Compton-Getting effect there is expected to be 
an anisotropy stemming from the diffusion of cosmic-rays in the 
interstellar medium. At high energies the main effects are
expected to be mainly due to the distribution of discrete CR sources and the
structure of the galactic magnetic field in the
neighborhood of the solar system. One study conducted consists of a simple
diffusion model assuming increased
production in the galactic disk due to
supernova remnants (SNR) \citep{PtSNR,SMP}. Also considered was
the diffusion of CR out of the galactic halo. This is attractive since we
have a deficit region in the general direction of the
North Galactic Pole which could be due in part to this diffusion. The main
contribution of CR from
SNR was considered for sources with
distances from Earth of $<1$ kpc and
ages $<0.05$ Myr. Calculations performed \citep{PtSNR,SMP} using these 
sources and taking into account CR re-acceleration as well as diffusion out of 
the galaxy gives an anisotropy
about 3 times greater than observed, with the main source of the
anisotropy due to the Vela SNR located at $128^{\circ}$ r.a.\ and
$-45.75^{\circ}$ dec. This model only predicts the magnitude of the expected
anisotropy, not its exact phase. It also is remarked that this
is a simplified model which assumes an isotropic diffusion tensor which is not
explicitly known to be true at these energies.

At energies of $\sim1$ TeV the heliosphere is 
believed to have an influence on the distribution of CR \citep{NgLC,NgTI}. One 
possible reason for the modulation of the anisotropy on the observed time scale 
could be due to variations in the heliosphere since we know that it changes 
in relation to solar output. It is noted 
that our data begins at the solar maximum and ends near the solar minimum. 
A recent derivation of the diffusion tensor contains a 
new component due to perpendicular spatial diffusion which is expected to be an 
important factor in understanding the anisotropy due to the Galactic disk 
as well as the modulation of CR in the outer heliosphere \citep{PerpD}. Finding 
a consistent explanation of the observed anisotropy 
and especially its time evolution will be a challenge.

\section{Acknowledgments}
We acknowledge Scott Delay and Michael Schneider for their dedicated efforts in the construction and maintenance of the Milagro experiment. This work has been supported by the National Science Foundation (under grants PHY-0245234, -0302000, -0400424, -0504201, -0601080, and ATM-0002744) the US Department of Energy (Office of High-Energy Physics and Office of Nuclear Physics), Los Alamos National Laboratory, the University of California, and the Institute of Geophysics and Planetary Physics.

\clearpage

\clearpage

\clearpage

\clearpage

\clearpage

\clearpage

\clearpage

\clearpage

\clearpage

\clearpage

\clearpage

\clearpage

\clearpage

\clearpage

%

\clearpage

\clearpage

\clearpage

\begin{deluxetable}{ccccccccc}
\tabletypesize{\footnotesize}
\tablecaption{Fit parameters to the sidereal anisotropy for all 18 declination bands.}\label{tab1}

\tablehead{
\colhead{Dec.} & \multicolumn{2}{c}{Fund.\ Harm.} & \multicolumn{2}{c}{1st Harm.} & \multicolumn{2}{c}{2nd Harm.} & \colhead{$\chi^2$/d.o.f.} & \colhead{Number}\\

\colhead{(mean)} & \colhead{Amplitude} & \colhead{Phase} & \colhead{Amplitude} & \colhead{Phase} & \colhead{Amplitude} & \colhead{Phase} & \colhead{} & \colhead{of events}\\
\colhead{} & \colhead{($\times10^{-3}$)} & \colhead{(deg)} & \colhead{($\times10^{-3}$)} & \colhead{(deg)} & \colhead{($\times10^{-3}$)} & \colhead{(deg)} & \colhead{} & \colhead{($\times10^{9}$)} 
}

\startdata

77.5&0.54$\pm$0.26&2$\pm$27&0.37$\pm$0.14&13$\pm$11&0.10$\pm$0.11&-39$\pm$19&262.57/234&0.65\\
72.5&0.73$\pm$0.14&22$\pm$11&0.19$\pm$0.08&-25$\pm$12&0.06$\pm$0.06&11$\pm$19&266.50/282&1.38\\
67.5&0.72$\pm$0.09&23$\pm$7&0.06$\pm$0.05&-24$\pm$26&0.01$\pm$0.04&28$\pm$116&308.67/330&2.39\\
62.5&0.83$\pm$0.07&19$\pm$5&0.12$\pm$0.04&-65$\pm$10&0.15$\pm$0.04&-2$\pm$4&355.61/330&3.63\\
57.5&0.99$\pm$0.06&6$\pm$3&0.12$\pm$0.03&-42$\pm$8&0.15$\pm$0.03&3$\pm$4&379.61/378&4.98\\
52.5&1.10$\pm$0.05&8$\pm$3&0.22$\pm$0.03&-60$\pm$4&0.17$\pm$0.03&6$\pm$3&406.42/378&6.31\\
47.5&1.31$\pm$0.04&8$\pm$2&0.33$\pm$0.03&-63$\pm$2&0.21$\pm$0.02&2$\pm$2&498.02/426&7.51\\
42.5&1.71$\pm$0.04&8$\pm$1&0.44$\pm$0.02&-68$\pm$2&0.26$\pm$0.02&1$\pm$2&475.85/426&8.46\\
37.5&1.95$\pm$0.04&6$\pm$1&0.45$\pm$0.02&-73$\pm$1&0.24$\pm$0.02&3$\pm$2&472.71/426&9.07\\
32.5&2.04$\pm$0.04&10$\pm$1&0.47$\pm$0.02&-76$\pm$1&0.20$\pm$0.02&1$\pm$2&520.47/426&9.28\\
27.5&2.17$\pm$0.04&9$\pm$1&0.53$\pm$0.02&-78$\pm$1&0.14$\pm$0.02&0$\pm$3&551.53/426&9.07\\
22.5&2.39$\pm$0.04&11$\pm$1&0.52$\pm$0.02&-81$\pm$1&0.12$\pm$0.02&-12$\pm$3&564.14/426&8.44\\
17.5&2.56$\pm$0.05&12$\pm$1&0.57$\pm$0.03&-81$\pm$1&0.10$\pm$0.02&-32$\pm$5&523.45/378&7.45\\
12.5&2.62$\pm$0.06&9$\pm$1&0.61$\pm$0.03&-85$\pm$2&0.05$\pm$0.03&-28$\pm$9&397.37/330&6.17\\
7.5&2.81$\pm$0.07&5$\pm$1&0.58$\pm$0.04&-80$\pm$2&0.08$\pm$0.03&-39$\pm$7&355.24/282&4.74\\
2.5&3.05$\pm$0.10&7$\pm$2&0.61$\pm$0.05&-80$\pm$3&0.14$\pm$0.04&54$\pm$5&280.00/234&3.31\\
-2.5&2.96$\pm$0.15&8$\pm$3&0.59$\pm$0.08&-89$\pm$4&0.23$\pm$0.06&56$\pm$5&271.14/186&2.04\\
-7.5&3.80$\pm$0.38&11$\pm$6&0.42$\pm$0.19&-85$\pm$13&0.14$\pm$0.13&52$\pm$18&118.02/90&1.04\\

\enddata
\tablecomments{All quoted errors are statistical and are used in the 
calculation of $\chi^2$.} 
\end{deluxetable}

\clearpage

\begin{deluxetable}{cccccccc}
\tabletypesize{\footnotesize}
\tablecaption{Fit parameters to the ST, UT and AST anisotropies using
data collected over all declinations.}\label{tab2}

\tablehead{
\colhead{Time} & \multicolumn{2}{c}{Fund.\ Harm.} & \multicolumn{2}{c}{1st Harm.} & \multicolumn{2}{c}{2nd Harm.} & \colhead{$\chi^2$/d.o.f.}\\

\colhead{Frame} & \colhead{Amplitude} & \colhead{Phase} & \colhead{Amplitude} &
 \colhead{Phase} & \colhead{Amplitude} & \colhead{Phase} & \colhead{}\\
\colhead{} & \colhead{($\times10^{-3}$)} & \colhead{(deg)} &
\colhead{($\times10^{-3}$)} & \colhead{(deg)} & \colhead{($\times10^{-3}$)} & \colhead{(deg)} & \colhead{}
}

\startdata

Sidereal&1.994$\pm$0.012&9.1$\pm$0.4&0.400$\pm$0.007&104.3$\pm$0.5&0.118$\pm$0.006&-0.9$\pm$1.0&936.3/426\\

Universal&0.365$\pm$0.012&217.4$\pm$1.9&0.043$\pm$0.007&168.7$\pm$4.8&0.029$\pm$0.006&93.2$\pm$4.3&520.9/426\\

Anti-Sidereal&0.120$\pm$0.012&47.1$\pm$5.8&0.019$\pm$0.007&155.5$\pm$11.1&0.013$\pm$0.006&4.9$\pm$9.4&425.9/426\\

\enddata
\tablecomments{All quoted errors are statistical and are used in the
calculation of $\chi^2$. For this single-dec-band analysis, the systematic errors exceed the statistical ones.}
\end{deluxetable}

\end{document}